\documentclass[fleqn,usenatbib]{mnras}

\usepackage[T1]{fontenc}
\usepackage{graphicx}	
\usepackage{amsmath}	
\usepackage{amssymb}	
\usepackage{latexsym}
\usepackage{array}
\usepackage{fancyhdr}
\usepackage[dvipsnames]{xcolor}
\usepackage[title]{appendix}
\usepackage{graphicx}
\usepackage{caption}
\usepackage{subcaption}
\usepackage{booktabs}
\usepackage[space]{grffile}
\usepackage{latexsym}
\usepackage{longtable}
\usepackage{xspace}
\usepackage{url}
\usepackage{hyperref}
\usepackage[utf8]{inputenc}
\usepackage[english]{babel}
\usepackage{pdflscape}
\usepackage{fancyhdr}
\usepackage{xcolor}
\usepackage{soul}
\usepackage{newtxtext,newtxmath}

\makeatletter

\newcommand{\hi}{{\sc H\,i}\xspace}

\newcommand{\hst}{\textit{HST}\xspace}

\newcommand{\spitzer}{\textit{Spitzer}\xspace}

\newcommand{\msun}{M$_\odot$\xspace}

\newcommand{\kms}{$\,$km$\,$s$^{-1}$\xspace}

\newcommand{\arcsecnew}{$\,$arcsec\xspace}
\newcommand{\arcsecsq}{$\,$arcsec$^{-2}\,$}
\let\oldsim\sim 
\renewcommand{\sim}{{\oldsim}}
\makeatother

\title{Neutral hydrogen lensing simulations in the Hubble Frontier Fields}
\author[Blecher et al.]{Tariq Blecher$^{1, 2}$, Roger Deane$^{3,4,1}$,  Danail Obreschkow$^{6, 7}$, Ian Heywood$^{5,1,2}$\\
$^1$ Centre for Radio Astronomy Techniques and Technologies, Department of Physics and Electronics, Rhodes University,-\\
Makhanda 6140, South Africa\\
$^2$ South African Radio Astronomical Observatory, 2 Fir Street, Observatory, 7925, South Africa \\
$^3$ Wits Centre for Astrophysics, University of the Witwatersrand, 1 Jan Smuts Avenue, 2000, Johannesburg, South Africa\\
$^4$ Department of Physics, University of Pretoria, Hatfield, Pretoria, 0028, South Africa \\
$^5$ Astrophysics, University of Oxford, Denys Wilkinson Building, Keble Road, Oxford, OX1 3RH, UK \\
$^6$ International Centre for Radio Astronomy Research (ICRAR), M468, University of Western Australia, WA 6009, Australia\\
$^7$ ARC Centre of Excellence for All Sky Astrophysics in 3 Dimensions (ASTRO 3D)
}
\date{Accepted XXX. Received YYY; in original form ZZZ}

\pubyear{2021}
\begin{document}
\label{firstpage}
\pagerange{\pageref{firstpage}--\pageref{lastpage}}
\maketitle

\begin{abstract}
Cold gas evolution ties the formation of dark matter halos to the star formation history of the universe. A primary component of cold gas, neutral atomic hydrogen (HI), can be traced by its 21-cm emission line. However, the faintness of this emission typically limits individual detections to low redshifts ($z\lesssim 0.2$). To address this limitation, we investigate the potential of targeting gravitationally lensed systems. Building on our prior galaxy-galaxy simulations, we have developed a ray-tracing code to simulate lensed HI images for known galaxies situated behind the massive Hubble Frontier Field galaxy clusters. Our findings reveal the existence of high HI mass, high HI magnification systems in these cluster lensing scenarios. Through simulations of hundreds of sources, we have identified compelling targets within the redshift range $z\approx 0.7 - 1.5$. The most promising candidate from our simulations is the  Great Arc at z=0.725 in Abell~370, which should be detectable by MeerKAT in approximately 50 hours. Importantly, the derived HI mass is predicted to be relatively insensitive to systematic uncertainties in the lensing model, and should be constrained within a factor of $\sim 2.5$ for a 95 per cent confidence interval.
\end{abstract}

\section{Introduction} 
Galaxy formation and evolution primarily involve gaseous flows and phase transitions. Gravitational accretion of gas onto dark haloes and radiative cooling within them \citep{white_1978} supplies galaxies with a reservoir of pristine neutral atomic material, primarily composed of hydrogen ($\sim75$ per cent by mass) and helium ($\sim25$ per cent). Provided that there are sufficient local instabilities and self-shielding, this gas further collapses into molecular clouds and stars, injecting thermal and mechanical feedback into the surrounding molecular and atomic interstellar medium \citep[ISM, e.g.][]{fierlinger_2016, hayward_2017}. A quantitative empirical understanding of this complex gas cycle and its cosmic evolution requires observations of atomic hydrogen (\hi) in large samples of galaxies over a wide range in redshift ($z$).

The new millennium has witnessed significant progress in extending the redshift range of large optical surveys to high $z$, well beyond the peak epoch of star formation at $z\sim2$ \citep[e.g.][]{lilly_2007,newman_2013, scodeggio_2018}. However, \hi is optically invisible, and its direct observation relies on a forbidden hyperfine transition corresponding to a radio line at 21 cm rest-frame wavelength (1.42~GHz). The weakness of this spectral line in emission has limited individual detections to the late-time universe ($z\lesssim 0.1$). There are only a few isolated emission detections reaching a few times further (e.g. \citealp{catinella_2018}), as well as stacking analyses \citep[e.g.][]{delhaize_2013,rhee_2013,bera_2019,chowdhury_2020}, intensity mapping \citep[e.g.][]{chang_2010, masui_2013}, and 21 cm absorption detections \citep[e.g.][]{gupta_2013,allison_2020} providing limited information out to $z\lesssim 3$.

A promising alternative to overcoming the inherent redshift limitations set by the weakness of the \hi emission line is strong gravitational lensing. 
Initial attempts to individually target several lensed \hi sources with short observing times have not yet resulted in clear detections \citep{hunt_2016, blecher_2019, ranchod_2022, chakraborty_2023}, however, these targets were selected on the basis of optical (not \hi) magnification estimates. Gravitational lensing occurs as the paths of light rays' paths are distorted in the presence of massive objects, magnifying distant objects by providing multiple lines of sight from observer to source. Can this phenomenon be leveraged to detect the faint neutral hydrogen 21-cm emission line in high-redshift galaxies? This question becomes increasingly relevant as next-generation cm-wave radio interferometers like the Square Kilometer Array (SKA) push back the \hi emission frontier to cosmological distances, which increases the likelihood of strong lensing occurences \citep[e.g.][]{deane_2015, deane_2016}. 

The total measured \hi flux of a lensed galaxy in units of JyHz, in the optically thin limit, is given by
\begin{equation}\label{eq:hi_flux}
 S_{\rm HI} = \frac{\mu_{\rm HI} M_{\rm HI}} {49.7 D_{\rm L}^2}\, , 
\end{equation}
 where $\mu_{\rm HI}$ is the average\footnote{More precisely, $\mu_{\rm HI}$ is the \hi mass-weighted magnification averaged over the source area.} \hi magnification, $M_{\rm HI}$ is the \hi mass in units of solar masses and $D_{\rm L}$ is the luminosity distance to the galaxy in units of megaparsec. Importantly, in this work, $M_{\rm HI}$ always refers to the {\it intrinsic or unlensed} \hi mass whereas $S_{\rm HI}$ always refers to the {\it apparent or lensed} \hi flux.

Cluster-scale lenses offer the highest magnifications over the largest angular scales and are natural targets for detecting multiple strongly-lensed systems within a relatively small area of sky \citep[e.g.][]{kneib_1993,kneib_1996,kneib_2011,oguri_2009,johnson_2014}. The most well-studied clusters from a lensing perspective are arguably the Hubble Frontier Fields \citep[HFF, ][]{lotz_2017}. The HFF campaign is a dedicated \textit{Spitzer Space Telescope} (\hst) and \textit{Spitzer Space Telescope} program to observe six of the most massive ($\gtrsim 10^{15}\, {\rm M_\odot}$) galaxy clusters at $z\approx 0.3-0.6$, which are favourable for optical-infrared (OIR) lensed sources at $z>6$.

The HFF campaign features a total of 840 \hst orbits and 1000 hours of \spitzer imaging, along with observations using numerous other instruments, including The Multi Unit Spectroscopic Explorer \citep[MUSE,][]{bacon_2010} on the Very Large Telescope (VLT). The photometric data spans a wide wavelength range from UV to near-infrared (0.2–8 $\mu$m), and there are hundreds of spectroscopic redshifts per field. With this dataset, the gravitational potential and associated lensing properties in these fields have been extensively modelled by many independent groups \citep[e.g][]{kawamata_2016,kawamata_2018,mahler_2018, lagattuta_2019}.

In this paper, we predict \hi emission magnifications, lensed images and fluxes to assess \hi magnification properties and detectability in the Hubble Frontier Fields, with a focus on known lensed sources identified in the literature.

The paper is structured as follows. In section~\ref{sec:hff_description}, we describe the fields, the lensing clusters, and the known background sources behind the cluster. In section~\ref{sec:sim_methods} we discuss the ray-tracing algorithm and chosen lens models used in the simulation. Section~\ref{sec:basic_stats} presents the results for the entire sample population of sources; and in section~\ref{sec:profiled_sources} we present a more detailed analysis of what we consider to be some of the most compelling individual targets. Finally, in section~\ref{sec:hff_discussion}, we explore the relationship between magnification and mass across the entire sample, estimate the observing time requirements, and investigate the effect of lensing systematics on our predictions for individual sources.

We assume a Planck 15 cosmology \citep{planckcollaborationandadam_2016} throughout, with $H_0 = 67.74\,{\rm km\,s^{-1}\,Mpc^{-1}}$, $\Omega_{\rm M} = 0.3075$ and $\Omega_{\Lambda} = 0.6910$.
\section{Description of the fields}\label{sec:hff_description} 
\subsection{Cluster lenses}
The primary cluster selection attribute for the HFF campaign was the probability of observing a $z=9.6$ galaxy magnified to 27~mag at 1.6\,$\mu$m \citep{lotz_2017}. This estimate was based on preliminary mass models, existing datasets \citep{postman_2012} as well as HST instrumental specifications

The six clusters chosen were Abell 2744 (A2744), MACSJ0416.1-2403 (M0416), MACSJ0717.5+3745, MACSJ1149.5+2223, Abell S1063 (AS1063), and Abell~370 (A370). We exclude the two clusters in the Northern Hemisphere (MACSJ0717.5+3745 and MACSJ1149.5+2223) from our study for two reasons. Firstly, these are not optimally observable by MeerKAT (along with its future successor, SKA1-mid), which is the key instrument on which we will focus to assess observational feasibility as it is the most sensitive interferometer in its class. Secondly,  these clusters are at a significantly higher redshift $z\approx 0.55$, and therefore they are less likely to  strongly lens \hi galaxies at $z\lesssim 1$ (lensing efficiency scales linearly with the angular diameter distance between lens and source). The central coordinates, as well as several key properties of the remaining clusters, are shown in Table~\ref{tab:hff_summary}. The selected clusters are in the redshift range $z\approx 0.3-0.4$, and all are extremely massive with virial masses $M_{\rm v} \gtrsim 10^{15}\, {\rm M_\odot}$. 

\begin{table*}
\caption{The {\bf upper sub-table} shows the key properties of each lensing cluster. The {\bf lower sub-table} shows the number of detections in the catalogue remaining after applying different selection criteria. The number of detections with spectroscopic redshifts are indicated in parentheses. For each source, we ensure that it: (i) has reliable photometry, (ii) is an extended source, (iii) situated within the boundaries of lensing model, (iv) is not associated with the lensing cluster, (v) has a reliable redshift, (vi) is outside the redshift range of the lensing cluster, and (vii) has an \hi flux above a minimum predicted cutoff. Refer to section~\ref{sec:backgrounds_srcs} for detailed information on the source selection criteria. The data were obtained from \citet{lotz_2017, shipley_2018}. \label{tab:hff_summary}}
\begin{tabular}{lllll}
Field                 & A2744    & AS1063    & A370    & M0416    \\
\toprule
{\bf Cluster Properties}  & & & & \\
\midrule
R.A. (J2000) (hms)     &   00 14 21.20 & 22 48 44.30 & 02 39 52.80 & 04 16 8.38 \\
Dec. (J2000) (dms)     &   -30 23 50.10 & -44 31 48.40 & -1 34 36.00 & -24 04 20.80 \\
Cluster Redshift  &    0.308    & 0.348    & 0.375    & 0.396    \\
No. Galaxies in Cluster &    79      & 90      & 75     & 49      \\
Virial Mass $/\,10^{15}$~M$_\odot$  & 1.8     & 1.4     & $\approx 1$  & 1.2     \\
\hline
{ \bf Number of identified OIR sources in catalogue} & & & & \\

\hline
All&		   9390 (546)&	7611 (237)&	6795 (221)&	7431 (389)\\
Criteria (i)-(iv)& 921 (151) &	643 (73)&	881 (85)&	742(140)\\
Criteria (i)-(vi)&             540 (151) &412 (37)&   543 (85)& 539 (140)\\
Final selection&		94 (18)		&	76 (20)			&	132 (37)		&	99 (37)	\\
\bottomrule

\end{tabular}
\end{table*}

\subsection{Known background sources}\label{sec:backgrounds_srcs} 
To perform predictions of lensed \hi for previously identified sources, we require a catalogue of known lensed galaxies. For this purpose, we rely on the public catalogue published in \citet{shipley_2018}. This comprehensive catalogue covers all the Frontier Fields and is based on photometric data spanning the UV to near-infrared (0.2–8 $\mu$m) wavelength range, complemented by a compilation of spectroscopic redshifts from the literature. 

As part of the data calibration process, cluster member galaxies and intra-cluster light (ICL) were modelled and subtracted before source finding and parameterisation. Due to limited spectroscopic coverage, photometric redshifts were calculated using a fit to the image spectral energy distribution (SED). This involved a linear combination of 12 galaxy templates, implemented by the {\sc eazy} code \citep{brammer_2008}. Stellar masses were estimated by the {\sc fast} \citep{kriek_2009} codebase, which fits stellar population synthesis templates to broadband photometry. Additionally the catalogue provides image-plane magnification estimates ($\mu_{\rm im}$) at the image centroid positions (i.e. peak flux position) for various lensing models.

For each entry in the catalogue, we predict an \hi mass from the apparent (i.e. magnified) stellar mass $M^{\rm im}_\star$ output by {\sc fast}. This calculation has two steps. First, we estimate the intrinsic stellar mass using the optical image plane centroid magnification $M_\star = M^{\rm im}_\star/\mu_{\rm im}$, utilising the latest available Clusters as Telescopes (CATS) model best $\mu_{\rm im}$ estimate \citep{mahler_2018,lagattuta_2019}. Secondly, we estimate $M_{\rm HI}$ using a $M_\star - M_{\rm HI}$ relation at $z=0$ \citep{maddox_2015}. We assume that the $M_\star - M_{\rm HI}$ relation does not evolve significantly out to the source redshifts considered, which is a conservative assumption \citep{sinigaglia_2022, chowdhury_2022, bera_2023}.

In the \citet{shipley_2018} catalogue, there are roughly $7000$ detections in each field identified by the source finder, including galaxies within the foreground cluster. We define a subset of this catalogue using several selection criteria, with resulting number counts shown in Table~\ref{tab:hff_summary}:
\begin{enumerate}
	\item The detection has reliable photometry (${\tt use\_phot\_flag}=1$).
	
	\item The photographic detection is extended and likely not a star (${\tt star\_flag}=0$).
	
	\item The photographic detection is within the boundaries of the deflection map used for ray-tracing simulations (see section~\ref{sec:lensmodel}).
	
	\item The photographic detection has not been identified as a galaxy within the lensing cluster (i.e. ${\tt source_{ID}}>20000$).
	
	\item If a spectroscopic redshift is not available then the photometric redshift has to be used. However, as photometric redshifts are significantly less reliable, a detection is only considered if the photometric redshift 68 per cent confidence interval is less than 20 per cent of its maximum value. For a Gaussian probability distribution, this is equivalent to the statement $\dfrac{\sigma_z}{\langle z \rangle} < 0.1$, i.e. that the standard deviation is less than 10 per cent of the expectation.
	
	\item The detection redshift $z_{\rm S}$ has to be larger than the cluster redshift $z_{\rm L}$ by a small margin $z > z_{\rm L}+0.04$. This is determined based on manual inspection of the cluster redshift distribution. This cutoff may not be sufficient to exclude all cluster galaxies; however, this is not important for this study as galaxies closer to the cluster will invariably have low magnifications and so would not contribute to the highly magnified statistics/sample.
	
	\item We filter out sources which are unlikely to be individually detected by MeerKAT within a practical observing time due to their extremely faint predicted flux. For this, we use a predicted \hi flux cut of $5$~JyHz which is equivalent to an unlensed, spatially-unresolved \hi galaxy with a mass of $M_{\rm HI} \approx 10^{9}\, {\rm M_\odot}$ at $z=0.4$. This flux cut is faint enough to leave substantial leeway for potentially higher \hi flux (due to either higher magnifications and/or mass) before sources are likely to become detectable by MeerKAT. The \hi flux is calculated using the mean predicted \hi mass and the optical image plane centroid magnification $\mu_{\rm im}$ from the latest available CATS model best estimate. Note that this filter limits the inclusion of high redshift sources, and there is no explicit upper redshift cutoff.
\end{enumerate}

These additional selection criteria reduce the number of candidates by roughly two orders of magnitude to a final count of 401 candidates. Unfortunately, the number of spectroscopic redshifts is limited, so we opt to use photometric redshifts for the majority of detections. Nevertheless, as will be seen, the best candidate sources usually have spectroscopic redshifts. To reduce the parameter space, we do not account for the remaining uncertainty on the photometric redshifts.

The source finding algorithm used in \citet{shipley_2018} identifies multiple images of the same galaxy as distinct entries in the detection catalogue. This  artificially increases the count of strongly magnified galaxies. A straightforward solution to this issue is to retain only the image with the largest stellar mass $M_\star$ in a multiply imaged system. We identify multiple images by cross-referencing catalogues in the literature \citep{kawamata_2016,kawamata_2018} as well as by ray-tracing the centroids of images to the source-plane and matching coordinates with small separations ($<50\,$kpc). In section~\ref{sec:profiled_sources}, when we delve into a more detailed analysis of the best candidate sources, we consider information from all images.

\begin{figure}
		\begin{center}
			\includegraphics[width=\columnwidth]{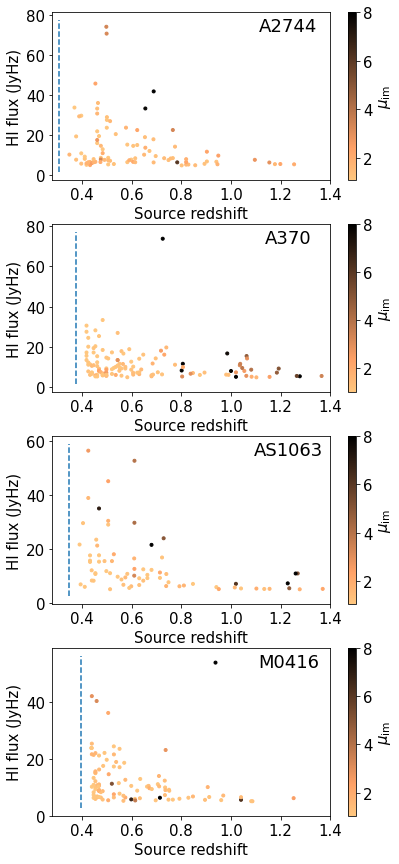}
			\caption{A naive \hi flux approximation is plotted against redshift for the lensed galaxy sample. In this approximation, the \hi magnification is set equal to the optical image plane centroid magnifications from the CATS v4.1 lensing model. The magnification factor is shown in colour scale and is saturated at $\mu_{\rm im} = 8$. The redshifts of the clusters are indicated by the dashed lines. \label{fig:hff_naiveflux}}
		\end{center}
\end{figure}

A naive \hi flux approximation against redshift for the lensed galaxy sample is presented in Figure~\ref{fig:hff_naiveflux}. In this approximation, the 
\hi magnification is set equal to the optical image plane centroid magnifications derived in \citet{shipley_2018} using the CATS v4.1 lensing model (see section~\ref{sec:lensmodel}). However, given the larger spatial extent of the \hi distribution, this point estimate is likely an over-estimate of the total \hi magnification. A primary aim of this paper is to provide a more accurate estimate of these lensed flux estimates using  sophisticated lens models in combination with empirically-derived \hi scaling relations. 

\section{Simulation methodology}\label{sec:sim_methods}
\subsection{Overview} %
We now describe the method used to predict the lensed \hi images, magnifications, and fluxes. Input to the pipeline is a lens model, in the form of the deflection angle map $\vec{\hat{\alpha}} (\vec{\theta})$, and a source catalogue, as described in section~\ref{sec:backgrounds_srcs}.

The basic components of a single simulation are: a parametric \hi radial distribution, a lens model, and a ray-tracing procedure. To model the \hi mass surface density $\Sigma_{\rm HI}$, we adopt the axisymmetric, thin-disk model of \citet{obreschkow_2009a},
\begin{equation}
\Sigma_{\rm HI} (r) = \frac{M_{\rm  H}/(2\pi r_{\rm disk}^2) \exp{(-r/r_{\rm disk})}}{1+R^{\rm c}_{\rm mol}\exp{(-1.6 r/r_{\rm disk})}},
\label{eq:HI_profile}
\end{equation}
where $r$ is the galactocentric radius in the disc plane, $M_{\rm H} = M_{\rm H_2}+M_{\rm HI}$, $r_{\rm disk}$ is the scale length of the neutral hydrogen disk (atomic plus molecular) and $R^{\rm c}_{\rm mol}$ is the amplitude of the exponential function describing the $M_{\rm H_2}/M_{\rm HI}$ ratio as a function of disk radius \citep{obreschkow_2009a}. 

The \hi mass is tightly correlated to the \hi size at $z\sim 0$ with a scatter of $\sigma \approx 0.06$~dex \citep{wang_2016}, described by,
\begin{equation}
\log_{\rm 10}(r_{\rm HI}/{\rm kpc}) = 0.506 \log_{\rm 10}(M_{\rm HI}/{\rm M_\odot}) - 3.293,
\label{eq:mass_size}
\end{equation}
where $r_{\rm HI}$ is defined as the diameter at which the \hi density drops to $\Sigma_{\rm HI} = 1~{\rm M_\odot\, pc^{-2}}$. Due to the tightness of the correlation, we expect that this relation is due to gas dynamics alone and therefore should hold to a higher redshift, however this is still to be verified.

\begin{figure}
    \centering
    \includegraphics[width=0.9\columnwidth]{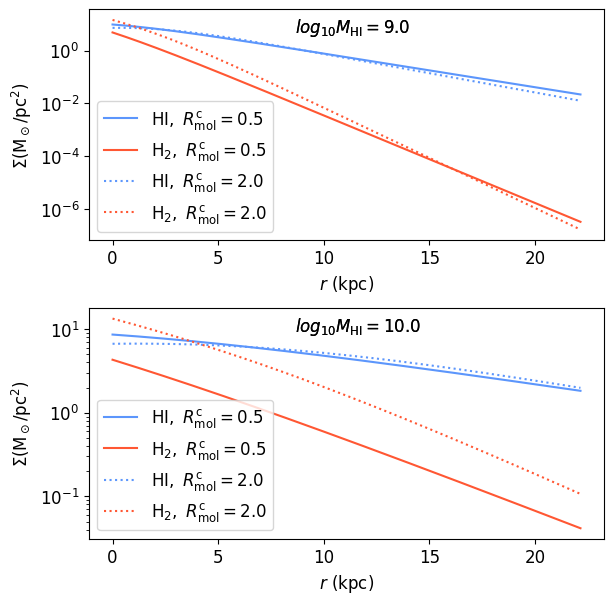}
    \caption{A suite of \hi radial density profiles demonstrating the theoretical models used to construct the \hi disks. The ${\rm H_2}$ radial density profile is shown as a comparision.}
    \label{fig:HI_radial_profiles}
\end{figure}

We calculate the value of $r_{\rm HI}$ using Equation \ref{eq:mass_size} and then use this to solve for $r_{\rm disk}$ in Equation \ref{eq:HI_profile} for an assumed $M_{\rm HI}$ and $R^{\rm c}_{\rm mol}$. We show several examples of these gas density profiles in Figure~\ref{fig:HI_radial_profiles}.

The lens models and the ray-tracing algorithm are discussed in section~\ref{sec:lensmodel} and appendix~\ref{sec:raytracing} respectively. As in \citet{blecher_2019}, we can marginalise over any nuisance parameters of the \hi disk model with an ensemble of simulations which sample the full parameter space. Our \hi lensing simulator is available at \url{https://github.com/TariqBlecher/tblenser}.

\subsection{Lens models}\label{sec:lensmodel} %
To model a lens, one solves for model parameters under the constraint that multiply imaged sources map to the same source-plane coordinate. The model complexity is also selected based on the signal-to-noise ratio (SNR) and angular resolution of the observation.

This is a challenging problem, especially for cluster lenses which have complex mass distributions. Regardless of the chosen model, a major difficulty in the optimisation are degeneracies in the parameter space, which limit the degree to which lens models can be constrained  \citep{meneghetti_2017,priewe_2017,acebron_2017,atek_2018}.

There are various approaches to lens model construction, roughly falling into two main classes of algorithms \citep{lefor_2013}. The first class are called parametric models \citep[e.g.][]{johnson_2014, mahler_2018}, which decompose the cluster mass distribution into physically-motivated analytic components, often using modified isothermal mass profiles. Parametric models typically assume "light traces mass" to approximate the mass profiles of cluster galaxies. The second class of models are called non-parametric (or grid-based), which use generic basis functions without a direct physical interpretation.

In section~\ref{sec:sim_results}, we use the models developed by the Clusters as Telescopes (CATS) and GLAFIC project teams. In section~\ref{sec:lens_systematics} we compare five different models to investigate systematic uncertainties. Both the CATS and GLAFIC teams use a parametric, light traces mass method. The CATS group uses the {\sc lenstool} \citep{jullo_2007,kneib_2011a} software package whereas the GLAFIC team uses the {\sc glafic} software \citep{oguri_2010}. Both sets of models fared well in comparison to other techniques when tested on synthetic data \citep{meneghetti_2017}, with {\sc glafic} achieving closest correspondence with the ground truth. The CATS models are updated frequently and hence have the advantage to use the latest available data. The public CATS models are over a field of view of $300-600$~arcsec at a resolution of $0.2-0.3$~arcsec whereas the GLAFIC models cover a smaller field of view ($160-180$~arcsec) at a higher resolution (0.03~arcsec). We use the CATS models in section~\ref{sec:basic_stats} as it covers all the candidate images and we use the GLAFIC models in section~\ref{sec:profiled_sources} for the more detailed profiled predictions which fall within the GLAFIC models' field of view due to its higher angular resolution. For lens models of each cluster, we opt for the latest models which are available on the public Mikulski Archive for Space Telescopes\footnote{https://archive.stsci.edu/pub/hlsp/frontier/}. For the CATS group, these are: Abell 2744 \citep{mahler_2018}, Abell~370 \citep{lagattuta_2019}, Abell S1063 v4.1 \citep{beauchesne_2023} and MACSJ0416.1-2403 v4.1. For the GLAFIC group, we use the latest models as given in \citet{kawamata_2016, kawamata_2018}, which correspond to the GLAFIC v4 model set.

\section{Simulation results}\label{sec:sim_results} 
\subsection{Detection and magnification statistics of full sample}\label{sec:basic_stats} %
In this subsection of simulation results, all detections in our refined catalogue are considered with the aim of identifying the most promising sources, which we then study individually in subsection~\ref{sec:profiled_sources}. To calculate magnification factors and \hi fluxes, we marginalise over any uncertainty in the \hi disk parameters. The disk position angle is sampled uniformly over the range $[0, 2\pi]$ radians, and the inclination angle $i$ is sampled from a sin$(i)$ distribution over the range $i \in [0,\pi/2]$ radians. The \hi mass is sampled from the log-normal distribution obtained from the $M_\star - M_{\rm HI}$ relation. Finally, we sample $R^{\rm c}_{\rm mol}$ from a log-normal distribution with $\mu_{\rm RC} = -0.1$ and $\sigma_{\rm RC} = 0.3$, consistent with the range of $M_{\rm H_2}/M_{\rm HI}$ quoted in \citet{catinella_2018} for the stellar mass range of these galaxies. 

\begin{figure*}
		\begin{center}
			\includegraphics[width=0.8\textwidth]{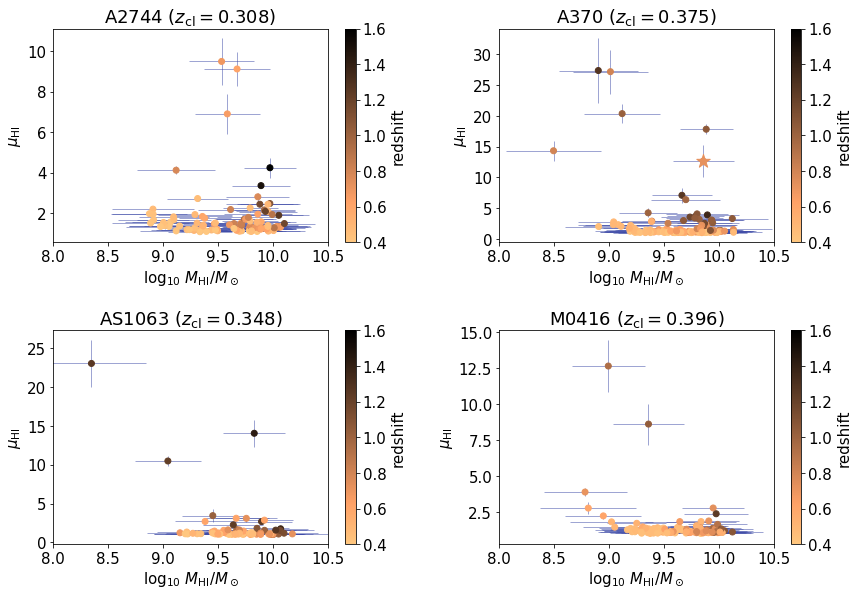}
			\caption{Total \hi magnification plotted against \hi mass for each field. The markers are coloured by redshift, and the Great Arc is indicated with a star symbol. For each data point, we marginalise over priors for the following variables: $R^{\rm c}_{\rm mol}$, inclination, position angle, and \hi mass. This plot highlights the possibility of high \hi mass systems with high \hi magnification within the context of cluster lensing.\label{fig:hff_magvmhi}}
		\end{center}
\end{figure*}

In Figure~\ref{fig:hff_magvmhi}, we plot the total \hi magnifications against \hi mass for the full galaxy sample, marginalising over the nuisance parameters. We observe high magnification systems across a broad range of \hi masses, even as high as $10^{10}\,$\msun despite the larger angular extent predicted by the $M_{\rm HI}$ - $D_{\rm HI}$ relation. All four clusters studied have strongly lensed HI galaxies, with Abell~370 having the greatest number of magnified candidates (at any magnification cutoff), and the most highly magnified systems included in our sample occur at redshifts $z\gtrsim 0.7$.

\begin{figure*}
	\begin{center}
			\includegraphics[width=0.8\textwidth]{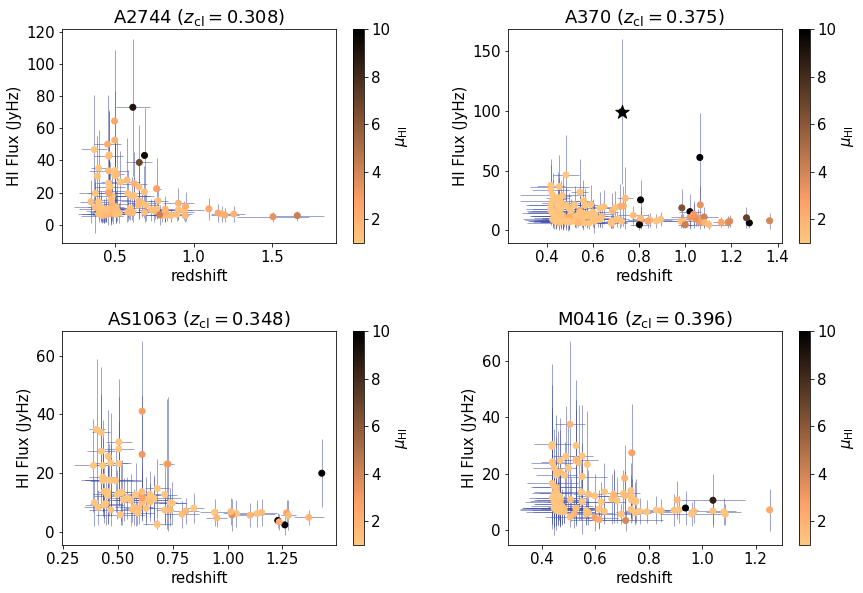}
			\caption{Integrated \hi flux plotted against redshift for each lensing cluster. The markers are coloured by \hi magnification and the colour saturates at $\mu_{\rm HI} = 10$. The Great Arc is marked with a star symbol. For each data point, we marginalise over priors for the following \hi disk variables: $R^{\rm c}_{\rm mol}$, inclination, position angle, and \hi mass. This plot provides a first order estimate of \hi detectability in the Frontier Fields and also conveys  contribution of magnification to the detectability.\label{fig:fluxvz}}
	\end{center}
\end{figure*}

In Figure~\ref{fig:fluxvz}, we show the \hi flux estimates as a function of redshift. The \hi flux distribution is computed by marginalising over probability distributions of the \hi disk parameters. This plot provides a first order estimate of source detectability and indicates the contribution due to magnification. We see that the higher flux sources within this sample at $z\ge 0.75$ are all strongly lensed. 

\subsection{Profiled sources}\label{sec:profiled_sources} 
We now focus on the most compelling candidates identified in our lensed galaxy sample. The highest \hi flux is predicted for the Great Arc at $z=0.725$ in Abell~370. Along with this source, we profile the $z=1.061$ triple image system in Abell~370, and the $z=1.429$ triple image system in Abell S1063. Further investigation of the highest \hi flux source in Abell 2744 ($z=0.61$) revealed a peculiar object without a spectroscopic redshift. This object is located near the edge of the \hst F814W image where the lens models are most uncertain due to fewer image constraints. It is adjacent to another object at a different photometric redshift and overlaps with masked pixels. Due to these complications, we decided against profiling this source.

Due to pixelisation and imperfections in the lens model, the centroids of the multiple optical images do not correspond exactly to the same point in the source-plane. The GLAFIC deflection maps have a higher angular resolution than the CATS maps, which in turn yields $\gtrsim 5$ times less scatter in the source-plane positions of multiple imaged galaxies. All our profiled sources are within the GLAFIC map field of view, and so we use the GLAFIC maps for these more detailed simulations. To address the source-plane centroid misalignment, we choose the source position that best reproduces the image-plane positions within the \hst-detected images. 
Note that the centroid scatter results in a negligible effect on the $\mu_{\rm HI}$, especially for the larger \hi disks. In addition to the source centroid position, each observed image will have a different stellar mass fit and therefore different predicted \hi mass and size (multiple images of the same galaxy are treated as multiple detections in the \citet{shipley_2018} catalogue). As in the previous subsection, we use the maximum stellar mass $M_\star$ of the different images as the best representation of the intrinsic stellar mass available. In the previous section, we assumed a uniform prior for the \hi disk position angle and a $\sin(i)$ prior on the disk inclination angle. These broad priors lead to a larger uncertainty in $\mu_{\rm HI}$. For certain sources, we are able to narrow these priors by ray tracing the lensed images as observed with \hst, and if the resulting source is a disk, we can estimate the inclination and position angle and fix these variables in the \hi disk model, under the assumption that the \hi broadly follows the stellar disk morphology.

For each profiled source, we conduct two experiments, where each experiment requires a different sampling of the \hi mass. The aim in the first experiment is to calculate $\mu_{\rm HI} (M_{\rm HI})$ (the dependence of the \hi mass on \hi magnification) without regard to the \hi mass expected from the $M_\star - M_{\rm HI}$ relation. It is purely an experiment to study the magnification properties without regard to detectability. To accomplish this, we sample $\log_{\rm 10}(M_{\rm HI})$ from a uniform mass distribution. This results in the $\mu_{\rm HI} (M_{\rm HI})$ curves shown in the left hand panel of Figure~\ref{profiled_source_stats}. These plots show how the average magnification of the sources change with the \hi mass and hence the angular scale of the source, where the individual plots will be discussed in more detail in following subsections.  

In the second experiment, our aim is to assess the \hi detectability. For this, we calculate the probability distribution of \hi flux ($S_{\rm HI}$) with $\log_{\rm 10}(M_{\rm HI})$ sampled from the Gaussian distribution obtained from the $M_\star - M_{\rm HI}$ relation. We run between 300 and 1000 simulations for each experiment, depending on the computational processing requirements. The results are shown in the right hand panel of Figure~\ref{profiled_source_stats}.

An alternative approach to studying the lensing properties is to use the source-plane magnification maps $\mu_{\rm src} (\vec{\beta})$ which represent the number of pixels in the image plane to which a source-plane pixel $\vec{\beta}$ is mapped (see Appendix~\ref{sec:raytracing} for further details). We construct these maps by generating the two-dimensional histogram of ray-traced image plane pixels, with each bin of the histogram corresponding to a source-plane pixel. This method shows the enhanced spatial resolution afforded by gravitational lensing and allows one to visualise the magnification profile of a given lensing system.

\subsubsection{The Great Arc in Abell~370 at z = 0.725}\label{sec:great_arc} %
\begin{figure*}
\subcaptionbox{An \hst filter-combined cutout of the Abell~370 Great Arc with image labels.\label{greatarc_labelled}}
{\includegraphics[width=0.95\columnwidth]{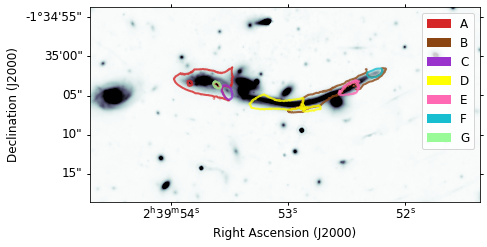}}\hfill
\subcaptionbox{An \hst filter-combined cutout of the Abell~370 Great Arc with the mean \hi image plane prediction shown with contour values set at $[0.6, 1]$~JyHz\arcsecsq \label{greatarc_detect}}
{\includegraphics[width=0.95\columnwidth]{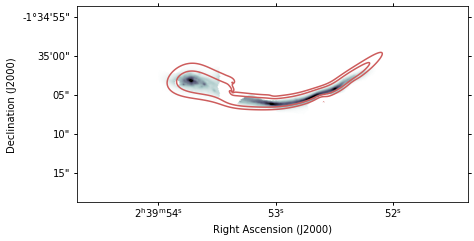}}
\subcaptionbox{An \hst filter-combined cutout of the Abell~370 triple with image labels.\label{a370triple_labelled}}
{\includegraphics[width=0.95\columnwidth]{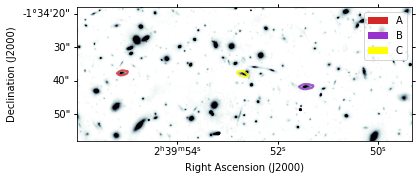}}\hfill
\subcaptionbox{An \hst filter-combined cutout of the Abell~370 triple with the mean \hi image plane prediction shown with contour values set at $[0.05,0.1,0.2]$~JyHz\arcsecsq.\label{a370tripledetect}}
{\includegraphics[width=0.95\columnwidth]{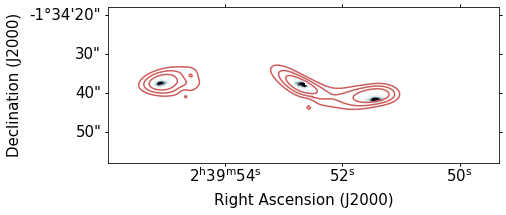}}
\subcaptionbox{An \hst filter-combined cutout of the Abell S1063 triple with image labels.\label{a1063triple_labelled}}
{\includegraphics[width=0.95\columnwidth]{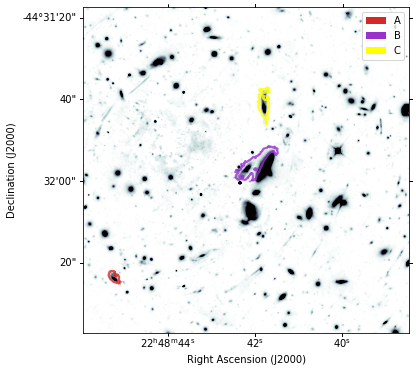}}\hfill
\subcaptionbox{An \hst filter-combined cutout of the Abell~370 triple with image labels with the mean \hi image plane prediction shown with contour values set at  $[0.05,0.1,0.2]$~JyHz\arcsecsq.\label{a1063tripdetect}}
{\includegraphics[width=0.95\columnwidth]{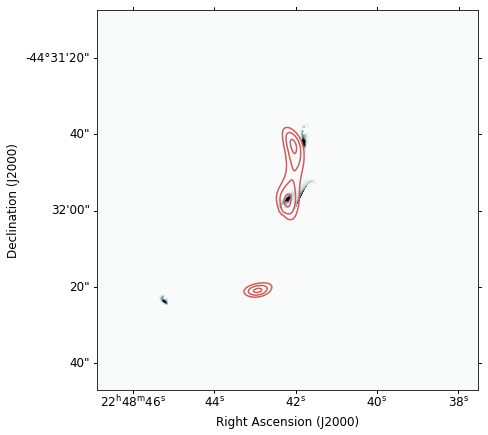}}
\caption{{\bf Left:} Cutouts of the \hst filter-combined detection images centred on the profiled source. We identify and label each image associated with the source in the input galaxy catalogue. {\bf Right:} Cutouts of \hst filter-combined detection images showing only the profiled sources. The mean \hi image plane prediction is shown with red contours. Optical data is taken from the \citet{shipley_2018} dataset}\label{image_plane_profiles}
\end{figure*}

The aptly named ``Great Arc" in Abell~370 spans over 20 arcseconds in the optical/infrared (see Figure~\ref{greatarc_labelled}) at a redshift of $z_{\rm spec}=0.725$ \citep{soucail_1988}. It consists of an image of a sheared disk galaxy, which is adjacent to the elongated main arc feature. Due to the peculiar extension and morphology of the arc, the source-finding procedure employed in \citet{shipley_2018} identified the Great Arc as 7 different images, labelled A-G in Figure~\ref{greatarc_labelled}. Summing over all images, the Great Arc has an apparent stellar mass of $\log_{\rm 10} (\mu M_\star/{\rm M_\odot}) = 11.5$.

\begin{figure*}
\centering
\subcaptionbox{The source-plane \hst filter-combined image of the Great Arc Image A and mean \hi source distribution,  with contour values set at $[0.8, 1, 1.5]$~JyHz\arcsecsq. \label{greatarc_sourceplane}}
{\includegraphics[width=0.75\columnwidth]{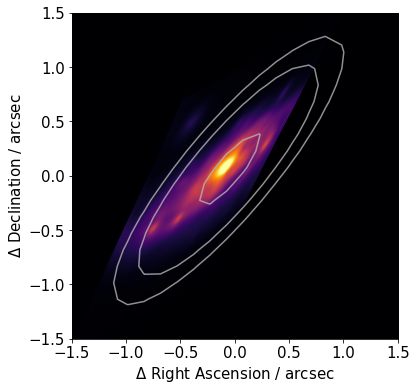}}\hfill
\subcaptionbox{The source-plane magnification map centred on the Great Arc source-plane centroid. The mean \hi disk is shown with contour values set at $[0.8, 1, 1.5]$~JyHz\arcsecsq.\label{greatarc_srcplanemag}}
{\includegraphics[width=0.9\columnwidth]{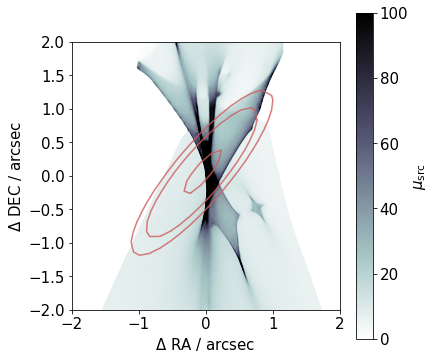}}
\subcaptionbox{The source-plane \hst filter-combined Image A of the Abell~370 triple and mean \hi source distribution is shown with contour values set at $[0.3,0.5]$~JyHz\arcsecsq.\label{a370trip_sourceplane}}
{\includegraphics[width=0.8\columnwidth]{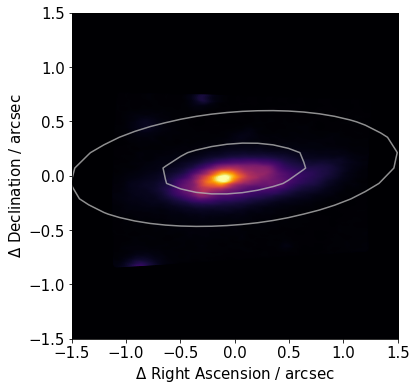}}\hfill
\subcaptionbox{The source-plane magnification map centred on Abell~370 triple image source-plane centroid. The mean \hi disk prediction is shown with contour values set at $[0.3, 0.5]$~JyHz\arcsecsq.\label{a370triple_srcplanemag}}
{\includegraphics[width=0.9\columnwidth]{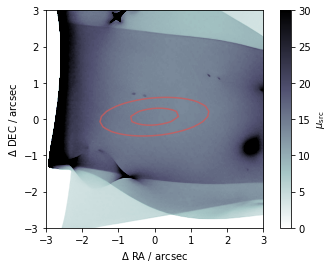}}
\subcaptionbox{The source-plane \hst filter-combined images for the Abell S1063 triple image ($z=1.429$). The left panel corresponds to Image A, the middle panel to Image B and the right panel to Image C.\label{a1063_source_panel}}
{\includegraphics[width=1.35\columnwidth]{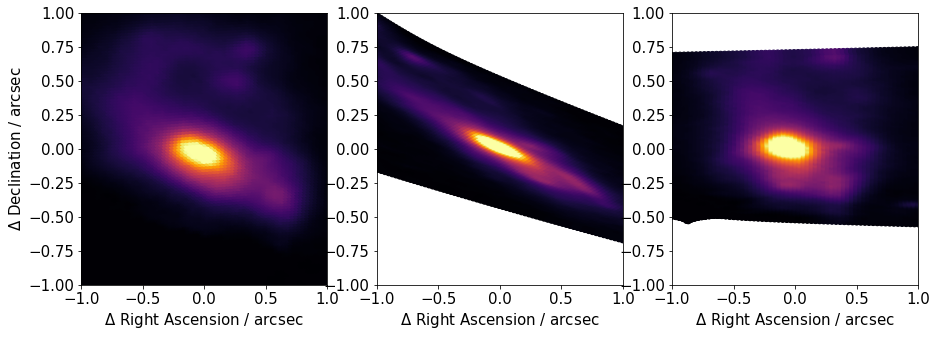}}\hfill
\subcaptionbox{The source-plane magnification of Image B in the Abell S1063 triple image ($z = 1.429$) is shown with a random realisation of the \hi source distribution shown with the contour values set at $[0.1,0.2,0.3]$~JyHz\arcsecsq.\label{a1063triple_srcplanemag}}
{\includegraphics[width=0.65\columnwidth]{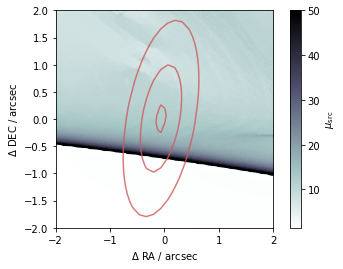}}
\caption{{\bf Left:} Selected source-plane \hst OIR filter-combined detection images in colour with gray \hi disk contour overlays. {\bf Right:} source-plane magnification maps in grayscale with an \hi source distribution shown with the red contour overlay. The source-plane magnification maps were smoothed with a 3-by-3 pixel median filter and upscaled by a factor of 4 using CNN-based Real-ESRGAN code \citep{wang_2021}}\label{profiled_sources_srcplane}
\end{figure*}

To better constrain the disk parameters, we ray-trace Image A into the source-plane (Figure~\ref{greatarc_sourceplane}). This results in a disk with a position angle of $\xi = 130 \pm 5$~degrees and an inclination angle of $i = 75 \pm 5$~degrees. This is consistent with \citet{richard_2010} who found a disk with projected major (minor) axes of 10.0~(2.5)~kpc (implying $i = 75.5$~degrees). In the same figure, we plot contours for the mean realisation of the source-plane \hi distribution. Considering the images of the Great Arc, Image A has the largest intrinsic stellar mass and is the least distorted representation of the original disk; therefore, we use its value in the catalogue for the \hi mass estimate and uncertainty, $\log_{\rm 10} (M_{\rm HI}/{\rm M_\odot}) = 9.84 \pm 0.27$.

When predicting \hi distributions from the input optical images, we find that the disk-like Image A does not exactly predict the observed shape of the arc and similarly, the images in the arc do not predict the central position of the disk-like image. This is due to a scatter ($\approx 0.2$\arcsec) in the source-plane centroid. To adequately fit both, we simply approximate the central coordinate in the source-plane as the average of the source-plane coordinates resulting from ray-tracing the centroid of Images A and C (Figure~\ref{greatarc_labelled}). The resulting average coordinate is $(\alpha, \delta) = $(02h\,39m\,53.34s, -01d\,34m\,48.28s). We find that this averaged source coordinate predicts the positions of both the arc and disk features.

In Figure~\ref{greatarc_detect}, we plot the \hst OIR image and overlay our mean predicted \hi distribution with the corresponding source-plane \hi distribution shown in Figure~\ref{greatarc_sourceplane}.

In Figure~\ref{greatarc_srcplanemag}, we show the source-plane magnification map for Abell~370 centred on the revised source coordinate of the Great Arc. The larger-scale structures correspond to cluster potential, and the smaller-scale structures correspond to individual lens galaxy potentials. As originally shown in \citet{richard_2010}, the central component of the source galaxy (shown in red contours) lies in the overlapping region between the two caustics and the western side of the disk lies in the cluster potential caustic.

\begin{figure*}
\centering
\subcaptionbox{Mass-magnification relation for the Great Arc in Abell~370. \label{greatarc_magvmhi}}
{\includegraphics[width=\columnwidth]{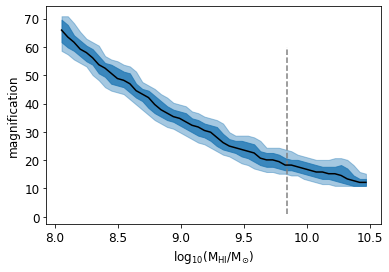}}\hfill
\subcaptionbox{Cumulative \hi flux probability distribution for the Great Arc in Abell~370. \label{greatarc_fluxprob}}
{\includegraphics[width=\columnwidth]{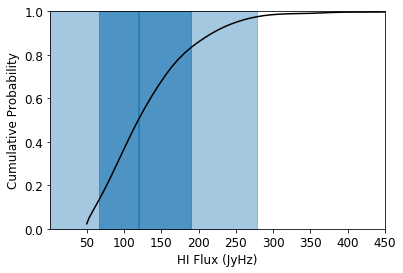}}
\subcaptionbox{Mass-magnification relation for the Abell~370 triple image.\label{a370trip_magvmhi}}
{\includegraphics[width=\columnwidth]{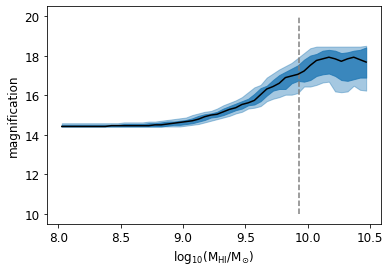}}\hfill
\subcaptionbox{Cumulative \hi flux probability distribution for the Abell~370 triple image.\label{a370trip_fluxprob}}
{\includegraphics[width=\columnwidth]{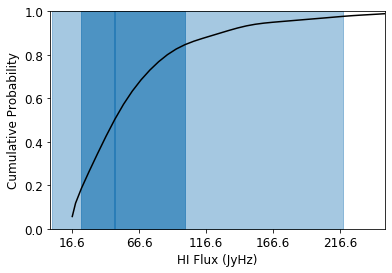}}
\subcaptionbox{Mass-magnification relation for the Abell S1063 triple image.\label{a1063trip_magvmhi}}
{\includegraphics[width=\columnwidth]{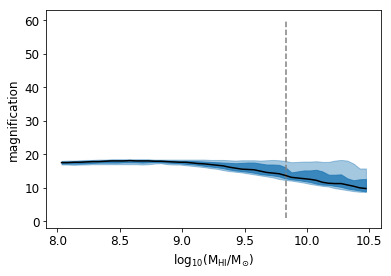}}\hfill
\subcaptionbox{Cumulative \hi flux probability distribution for the Abell S1063 triple image.\label{a1063trip_fluxprob}}
{\includegraphics[width=\columnwidth]{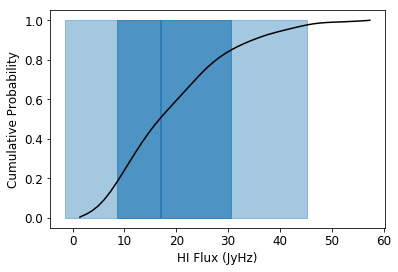}}
\caption{{\bf Left:} Total \hi magnification as a function of total \hi mass for the profiled sources. The black curve shows the mean expectation, while the dark and light blue filled areas show the 68 and 95 per cent confidence intervals respectively. The gray dashed line shows the \hi mass prediction based on the stellar mass. {\bf Right:} The cumulative flux probability is shown with the black curve, while the 68 and 95 per cent confidence intervals of the \hi flux distribution are shown by the dark and light blue filled areas respectively.\label{profiled_source_stats}}
\end{figure*}

The $\mu_{\rm HI} (M_{\rm HI})$ profile for the Great Arc is shown in Figure~\ref{greatarc_magvmhi}. We find a monotonically decreasing function moving from $\mu_{\rm HI} \approx 65$ at the low mass end to $\mu_{\rm HI} \approx 12$ at the high mass end. The overlay with the mean source-plane \hi disk allows one to visualise the effect of differential magnification as well as how variation in the size of the disk would change the magnification, i.e. smaller disks have larger fractions of their mass in high magnification regions.

Marginalising over the predicted mass distribution, we find a best estimate of $\mu_{\rm HI} = 19 \pm 4$, which is similar to the image plane magnifcation averaged over all images $\mu_{\rm im} = 22$. The \hi flux probability is shown in the lower panel, where we find an estimated \hi flux of $S_{\rm HI} =119^{+70}_{-52}$~JyHz.



\subsubsection{Triple image in Abell~370 at z = 1.061} %
In section~\ref{sec:basic_stats}, we identified a high redshift triple image with a spectroscopic redshift of $z=1.061$, which can be seen in the \hst filter-combined image (Figure~\ref{a370triple_labelled}). The centroids of the two outer images are spaced approximately $37$~arcsec apart. Summing over the three images, yields an apparent stellar mass of $\log_{\rm 10} (\mu M_\star/{\rm M_\odot}) = 11.4$.

Ray tracing the images to the source plane reveals an inclined disk galaxy (Figure~\ref{a370trip_sourceplane}), with an inclination of $i = 70 \pm 8$~degrees and a position angle of $\xi = 175 \pm 10$~degrees. We find that the coordinates of the three images are best reproduced when the average of the three source-plane centroid positions is used as input. This averaged source-plane centroid, $ (\alpha, \delta) = {\rm (02h\, 39m\, 53.51s -01d\, 34m\, 36.87s)}$, is used in these simulations. For the \hi mass and associated uncertainty, each image yields an almost identical prediction, with the maximum $M_{\rm HI}$ prediction of the three images being $\log_{\rm 10} (M_{\rm HI}/{\rm M_\odot}) = 9.91 \pm 0.28$. 

We show the mean \hi image prediction (Figure~\ref{a370tripledetect}) overlaid on a cut-out of the filter-combined \hst image. Due to the extension of the \hi distribution, the \hi images of B and C overlap to form an arc, connecting the two optical images.

The dependence of \hi mass on magnification is shown in Figure~\ref{a370trip_magvmhi}. Insight into the $\mu_{\rm HI} (M_{\rm HI})$ profile can be obtained from the source-plane magnification map shown in Figure~\ref{a370triple_srcplanemag}. The galaxy appears to be situated in a remarkably uniform, extended, high magnification region near one of the cluster caustics, as well as overlapping with a smaller scale mass distribution. An increase in the extension of the source distribution would move more of the distribution onto the maximal ($\mu_{\rm src} > 25$) magnification portion of the caustic.

We estimate an \hi flux of $S_{\rm HI} =48^{+52}_{-25}$~JyHz and a total \hi magnification of $\mu_{\rm HI} = 17 \pm 1$ which is much greater than the point image-plane estimate, $\mu_{\rm im} = 4.8$, used in Figure~\ref{fig:hff_naiveflux}.

\subsubsection{Triple image in Abell S1063 at z = 1.429} %
This triple image system at a spectroscopic redshift of $z=1.429$ \citep{balestra_2013,richard_2014,johnson_2014} is predicted to exhibit the highest \hi flux for redshifts $z\gtrsim 1.1$ (Figure~\ref{fig:fluxvz}). Figure~\ref{a1063triple_labelled} shows the image positions on a cutout of the \hst filter-combined image of the Abell S1063 field. The two outer images are separated by approximately 55\arcsecnew. Summing over the three images, yields an apparent stellar mass of $\log_{\rm 10} (\mu M_\star/{\rm M_\odot}) = 10.8$.

When each image is ray-traced to the source-plane (Figure~\ref{a1063_source_panel}), we encounter significant ambiguity and inconsistency in the shape and orientation of the source galaxy between the three images. Since we cannot securely constrain the inclination and position angles, we instead sample these from uniform distributions, as done in section~\ref{sec:basic_stats}. In addition, the source-plane positions of the three images could not be reconciled straightforwardly. For the source-plane centroid, we take the ray-traced position of the middle image, B. In practice, using the centroids of images B and C produces similar images and magnifications, whereas the centroid of image A fails to reproduce the images B and C. The disadvantage of using this approach is that the prediction of image A is in an incorrect position, as shown in Figure~\ref{a1063tripdetect}. We note that this inconsistency may also differ between lens models. The maximum $M_{\rm HI}$ prediction of the three images being $\log_{\rm 10} (M_{\rm HI}/{\rm M_\odot}) = 9.83 \pm 0.27$. 

The dependence of mass with magnification and the probability distribution of the \hi flux is shown in Figure~\ref{a1063trip_magvmhi}. We estimate an \hi flux of $S_{\rm HI} =17^{+13}_{-8}$~JyHz and an \hi magnification of $\mu_{\rm HI} = 14 \pm 2$ which is significantly larger than the image plane magnification estimates at the image centroids, $\mu_{\rm im} \approx 3-7$. The mass-magnification profile is of a different class than what was seen previously, being largely flat and decreasing slightly at the high mass end. Insight into the $\mu_{\rm HI} (M_{\rm HI})$ profile can be obtained from the source-plane magnification map, which is shown in  Figure~\ref{a1063triple_srcplanemag}. The galaxy appears to overlap with one of the cluster caustics and partly falls within an extended high magnification region.

\section{Discussion}\label{sec:hff_discussion} 
\subsection{HI magnification properties}\label{sec:mag_properties}%
\begin{figure}
	\centering
	\includegraphics[width=\columnwidth]{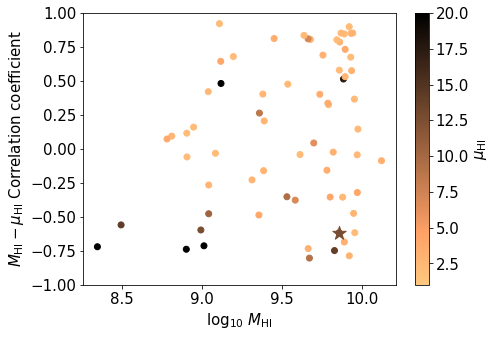}
	\caption{The Pearson correlation coefficient ${\rm corr(\mu_{\rm HI}, M_{\rm HI})}$ for the strong lensed ($\mu_{\rm HI}>2$) \hi sources in all four fields. Values with positive correlation mean that $\mu_{\rm HI}$ tends to increase with $M_{\rm HI}$ (after marginalising over the other disk parameters) while negative correlation values indicate the reverse. The Great Arc is indicated with a star marker.\label{magvmhi_corr}}
\end{figure}

In \citet{blecher_2019}, we showed that for low redshift ($z\lesssim0.4$), galaxy-galaxy lensing systems with arcsecond-scale Einstein radii and small impact factors, the magnification of each galaxy was a monotonically decreasing function of \hi mass (i.e. a negative correlation). We now explore $\mu_{\rm HI} (M_{\rm HI})$ for lensed sources in the Hubble Frontier Fields clusters which have more complex mass density profiles compared to galaxy-scale lenses. In Figure~\ref{magvmhi_corr}, we present the correlation coefficients, ${\rm corr} (\mu_{\rm HI},M_{\rm HI})$, for all strongly lensed sources ($\mu_{\rm HI} > 2$) in the fields studied. We observe a more complex picture than in the galaxy-galaxy lensing case, with 37 out of 67 strongly lensed sources having positive correlation. However, for a higher magnification cut $\langle\mu_{\rm HI}\rangle > 10$, only 2 out of 10 sources have positive correlations.

For the Great Arc, even though the magnification is a monotonically decreasing function of \hi mass, the magnification remains high ($\mu_{\rm HI} > 10$) for all \hi masses, with our best estimate $\mu_{\rm HI} = 19 \pm 4$. The Abell~370 triple image at $z = 1.061$ exhibits an interesting $\mu_{\rm HI} (M_{\rm HI})$ profile, which is constant in the lower and higher mass ranges but increases in the $9<\log_{\rm 10} (M_{\rm HI}/{\rm M_\odot})<10$ range, with a best estimate of $\mu_{\rm HI} = 17 \pm 1$. The Abell S1063 triple image at $z = 1.429$ has a constant $\mu_{\rm HI} (M_{\rm HI})$ profile at $\mu_{\rm HI} \sim 18$ until $\log_{\rm 10} (M_{\rm HI}/{\rm M_\odot}) \approx 9$, after which it declines, with a best estimate of $\mu_{\rm HI} = 14 \pm 2$. 
\subsection{Detection prospects}\label{sec:HFF_detectprospects} %
We now estimate the observing time required to detect the profiled lensed sources using the MeerKAT telescope. The estimation is based on a frequency-integrated $5\sigma$ detection with telescope sensitivity and imaging parameters listed in Table~\ref{tab:HFF_detectionprospects}. 

To calculate a realistic observing time $\tau$ requirement, we use the following equation,
\begin{equation}
\tau_{\rm obs} = \left(\frac{R_{\rm S/N} S_{\rm SEFD}}{S_{\rm HI}} \frac{w_{\rm nat}}{w_{\rm Briggs_{\tt robust=0.5}}}\right)^2 \frac{\mathrm{d}\nu}{2N_{\rm a}(N_{\rm a}-1)}\left(1 +\frac{A_{\rm galaxy}}{A_{\rm beam}}\right),
\end{equation}
where $R_{\rm S/N}$ is the required signal-to-noise ratio; $S_{\rm SEFD}$ is the system equivalent flux density per antenna in units of Jy; $\mathrm{d}\nu$ is the line width in units of Hz; $N_{\rm a}$ is the number of antennas in the array; $(1 +\frac{A_{\rm galaxy}}{A_{\rm beam}})$ accounts for the source flux being distributed over multiple beams \citep{meyer_2017}; and $\frac{w_{\rm nat}}{w_{\rm Briggs_{\tt robust=0.5}}}$ describes the change in sensitivity due to a Briggs robust$=0.5$ imaging weighting. Although natural weighting would be ideal for maximising the signal of an unresolved source, natural weighting also has the largest PSF sidelobes and therefore can result in lower fidelity due to noise artefacts remaining after deconvolution. We calculate a $(u,v)$ weighting related decrease in sensitivity of $w_{\rm Briggs 0.5}/w_{\rm nat} =0.8$ by imaging a simulated MeerKAT dataset with only Gaussian noise at the two weightings.

We assume sources have a velocity width of $200$~\kms{} and that 60 antennas participate in the observation. To calculate the beam size, we simulate an observation and use the {\sc wsclean} \citep{offringa_2014} fitted-beam values. We estimate the galaxy area as the image area with flux above 1 per cent of its peak value.

With MeerKAT, we find that the Great Arc has a mean $5\sigma$ detection time of 16~hr and a 68 per cent confidence interval upper limit of 51~hr. The Abell~370 triple image has a 68 per cent confidence interval lower limit of 30~hr and hence observation of this target could be commensal with a Great Arc observation. The Abell S1063 triple image at $z = 1.429$ would require an unreasonable observation time with MeerKAT ($\gg 200$~hours). 

The large uncertainties in the detection time estimates are primarily driven by uncertainties in the \hi mass. In future, this may be improved by using other $M_{\rm HI}$ estimators, such as the stellar mass density \citep{catinella_2018}, or angular momentum \citep{obreschkow_2016}.

\begin{table*}
\caption{On-source observing time estimates for the profiled sources with the MeerKAT telescope. MeerKAT technical specifications were taken from SARAO observatory reports and usage experience. See text for further details.  \label{tab:HFF_detectionprospects}}
\begin{tabular}{|lccccccr|}
\toprule
Source& z & $\nu_{\rm obs}$ &$ S_{\rm HI}$ &$S_{\rm SEFD}$& $A_{\rm beam}$ & $A_{\rm galaxy}$& $\tau_{\rm obs} (5\sigma$)\\
Unit& & (MHz)& (JyHz)&  $({\rm Jy})$ &(arcsec$^2$)&(arcsec$^2$)&(hr)\\
\midrule
A370 Great Arc&0.725 &823& $119^{+70}_{-52}$  & 475.0& 1586 & 358 &$16^{+35}_{-10}$\\[0.2cm]
A370 triple   &1.061 &689& $48^{+52}_{-25}$   & 550.0& 1950    & 783  &$131^{+441}_{-101}$\\[0.2cm]
A1063 triple  &1.429 &584& $17^{+13}_{-8}$    & 620.0&  2225   & 494 &$974^{+2878}_{-670}$\\
\bottomrule                       
\end{tabular}
\end{table*}

\subsection{HI mass reconstruction accuracy}
We now assess the accuracy with which $M_{\rm HI}$ can be reconstructed assuming the observed source is described by the analytic \hi source model defined in section~\ref{sec:sim_methods}. 

First, we compute $S_{\rm HI}$ (Equation~\ref{eq:hi_flux}) for the simulations with logarithmic priors $p(M_{\rm HI}) \sim 1/M_{\rm HI}$ (i.e. $p({\rm log} (M_{\rm HI}))$ is constant). Note that the distribution over $\mu_{\rm HI}$ is now also represented in the distribution over $S_{\rm HI}$.

We can then estimate the conditional probability distribution,
\begin{equation}\label{eq:mhi_conditional}
p(M_{\rm HI}|S_{\rm HI}) \propto p(S_{\rm HI}|M_{\rm HI}) p(M_{\rm HI}).
\end{equation}

\begin{figure}
	\begin{subfigure}{0.49\textwidth}
			\centering
		\includegraphics[width=\textwidth]{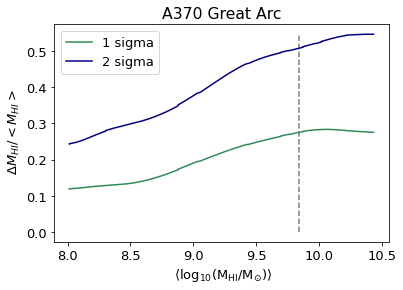}
	\end{subfigure}
 \begin{subfigure}{0.49\textwidth}
		\centering
	\includegraphics[width=\textwidth]{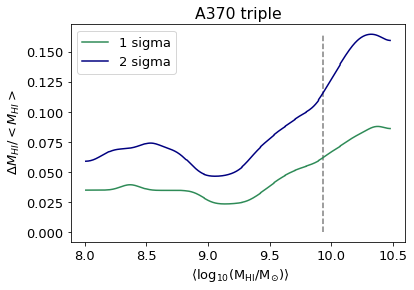}
\end{subfigure}
\begin{subfigure}{0.49\textwidth}
		\centering
	\includegraphics[width=\textwidth]{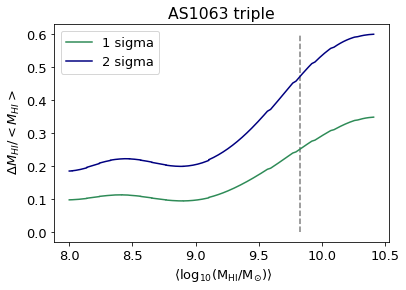}
\end{subfigure}
\caption{Relative uncertainties in the \hi mass reconstruction are shown as 68 and 95 per cent confidence intervals normalised by the expectation of the \hi mass. The dashed vertical lines represent the expectation of the \hi mass from the stellar mass conversion. Note that these uncertainties arise solely from the lens modelling only and do not consider measurement noise. \label{fig:mhi_recon}}
\end{figure}

Using Equation~\ref{eq:mhi_conditional}, we calculate relative uncertainties $\Delta M_{\rm HI}/\langle M_{\rm HI}\rangle$ as a function of $S_{\rm HI}$, where $\langle M_{\rm HI} \rangle$ is the expectation of $p(M_{\rm HI}|S_{\rm HI})$ and $\Delta M_{\rm HI}$ represents a confidence interval of $p(M_{\rm HI}|S_{\rm HI})$. The results are shown in Figure~\ref{fig:mhi_recon}. We observe that, for a (68, 95) per cent confidence interval, the relative uncertainty on the \hi masses at the best estimate of the predicted masses are approximately: (26, 51) per cent for the Great Arc; (6, 11) per cent for the A370 triple; and (25, 47) per cent for the AS1063 triple. 

To include the effect of measurement noise,  we use Bayes Theorem to infer the probability distribution of the \hi flux,
\begin{equation}
p(S_{\rm HI}|S_{\rm 0}) \propto p(S_{\rm 0}|S_{\rm HI})p(S_{\rm HI}),     
\end{equation}
where $S_{\rm 0}$ is the measured flux value.
The prior can be set to ensure positivity, and the likelihood, under the assumption of Gaussian noise, becomes 
\begin{equation}
p(S_{\rm 0}|S_{\rm HI}) = \frac{1}{\sigma_S\sqrt{2\pi}} \exp{\left[-\frac{(S_{\rm 0}-S_{\rm HI})^2}{2\sigma_S^2}\right]},
\end{equation}
where an independent estimate of $\sigma_{S}$ can be obtained the spectral cube.  

We can then marginalise over the intermediary $S_{\rm HI}$ to obtain the posterior of the \hi mass \footnote{We note that this reconstruction method, employing the full PDF $p(\mu_{\rm HI} | M_{\rm HI})$, differs from the method used in \citet{blecher_2019} which only used $\langle \mu \rangle (M_{\rm HI})$.},
\begin{equation}
p(M_{\rm HI}|S_{\rm 0}) \propto \int  p(M_{\rm HI}|S_{\rm HI}) p(S_{\rm HI}| S_{\rm 0}) {\rm d} S_{\rm HI}.
\end{equation}

We now re-compute the relative uncertainties including noise, for a measurement of $S_0 \approx S_{\rm HI}$ with 5~$\sigma$ noise level. We find that the relative uncertainties at (68, 95) per cent confidence intervals are now (65, 136) per cent for the Great Arc; (40, 82) per cent for the A370 triple; and (59, 123) per cent for the A1063 triple. In summary, we find that the relative uncertainty has increased by (30-40, 70-80) per cent for the (68, 95) per cent confidence intervals when noise is included. In summary, the \hi mass of all three sources can be constrained to within a factor of $\approx 2.5$ for a $5~\sigma$ measurement within a 95 per cent confidence interval.

We now assess potential bias from a more general flux measurement (i.e. not restricting $S_0 \approx S_{\rm HI}$). We conduct a hypothetical experiment with a single ground truth \hi mass, magnification, and flux value $S_{\rm HI}$ for each profiled source as outlined in Table~\ref{tab:HFF_recon}. We sample 1000 measured flux values from the normal distribution $S_{\rm 0} \sim \mathcal{N}(S_{\rm HI},\sigma_{S})$, where each sample represents a different realisation of the measurement noise, and $\sigma_{S} = S_{\rm HI} / 5$, which results in a 5~$\sigma$ observation on average. 

The average results of this approach are displayed in the rightmost column of Table~\ref{tab:HFF_recon}. On average, the reconstructed mass using the full flux distribution is consistent with the true \hi mass. For individual realisations, we find that the statistics approximately follow Guassian distribution with $\sim 65$ per cent of realisations within a 68 per cent confidence interval and $\sim 93$ per cent of realisations within a 95 per cent confidence interval of the true \hi mass. 

\begin{table*}
\caption{Data for the hypothetical \hi mass reconstruction experiment. The first three columns indicate the ground truth \hi mass, magnification and flux values. The last column shows the reconstructed \hi masses averaged over 1000 realisations of the observational noise. \label{tab:HFF_recon}}
\begin{tabular}{|lcccr|}
\toprule
Source& $M_{\rm HI}$ & $\mu_{\rm HI}$ & $S_{\rm HI}$& $M_{\rm recovered}$\\
\midrule
A370 Great Arc& 9.84& 17.37 & 116.5&$9.73^{+0.14}_{-0.18}$\\[0.2cm]
A370 triple   & 9.91& 17.01 & 53.0&$9.85^{+0.09}_{-0.13}$\\[0.2cm]
A1063 triple  & 9.83& 16.70 & 20.4&$9.83^{+0.13}_{-0.15}$\\
\bottomrule                       
\end{tabular}
\end{table*}

Hence, given an \hi flux measurement, for these three sources, $M_{\rm HI}$ should be well constrained  under the assumptions that the \hi disks are adequately represented by an axisymmetric disk with a smooth, double-exponential radial density profile, and that the $z=0$ $M_{\rm HI}\,-\,D_{\rm HI}$ correlation holds at higher redshifts. Future work could test this analytic approach by ray-tracing observed \hi galaxy profiles, such as those from the THINGS sample \citep{leroy_2008}, or realistic cubes from hydrodynamical simulations \citep[e.g.][]{pillepich_2018, dave_2019}, validating whether the recovered \hi masses are reliable. 

\subsection{Lens model uncertainty}\label{sec:lens_systematics}
To model a cluster lens, one has to optimise over the parameters describing the cluster components. Parameter uncertainties in the lens model can be estimated with a Markov Chain Monte Carlo algorithm \citep[MCMC;][]{jullo_2007,kawamata_2016}, nested sampling \citep{beauchesne_2023} or other techniques. However, this may not fully account for the systematic uncertainty associated with the underlying model assumptions \citep[e.g.][]{limousin_2016}. Systematic errors on the deflection angles can arise from light-of-sight projection effects \citep{meneghetti_2010}, scatter in mass-to-light scaling relations \citep{daloisio_2011}, uncertainties in the cosmological model \citep{bayliss_2015} and unmodelled structures along the line of sight \citep{host_2012}. To estimate the magnitude and impact of systematic uncertainties, the variance between multiple lens models can be used \citep[e.g.][]{atek_2018}.

We recreate the \hi mass-magnification profile for the two candidates for which the disk inclination and position angle could be constrained (i.e. the Great Arc in Abell~370 and the $z=1.061$ triple image in Abell~370) using five independent mass models. We use all available maps on the STScI public repository which were based on \hst data, and have an image resolution of $\lesssim 0.2$~arcsec (GLAFIC \citep{oguri_2010, kawamata_2016, kawamata_2018}, CATS \citep[e.g.][]{mahler_2018,lagattuta_2019}, Keeton \citep[e.g.][]{ammons_2014, mccully_2014}, Sharon \citep{johnson_2014} and Williams \citep[e.g.][]{jauzac_2014,grillo_2015}). All teams have used parametric methods except for the Williams team.

\begin{figure}
	\begin{subfigure}{0.49\textwidth}
			\centering
		\includegraphics[width=\textwidth]{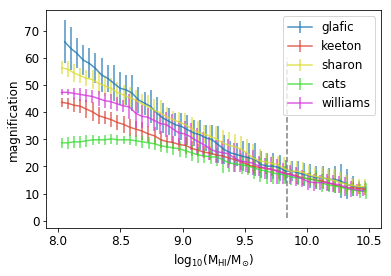}
	\end{subfigure}
\begin{subfigure}{0.49\textwidth}
		\centering
	\includegraphics[width=\textwidth]{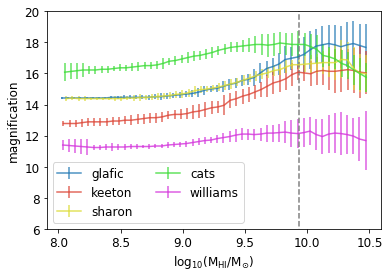}
\end{subfigure}
	\caption{\hi mass-magnification profiles of the Great Arc in Abell~370 (upper) and the $z=1.061$ triple image in Abell~370 (lower) using multiple lens models. Each curve shows the mean expectation, with the error bars denoting 68 percent confidence interval. The gray dashed line shows the \hi mass predictions based on the stellar mass.\label{fig:systematic}}
\end{figure}

For low \hi masses $M_\mathrm{HI} \lesssim 10^9\,$\msun, the magnification estimates are not consistent within $1\sigma$ (see Figure~\ref{fig:systematic}). However, for higher \hi masses and hence more extended \hi distributions, the magnifications predicted by the different models are within $1\sigma$ (except for the Williams model in the case of the triple image). Our model predicts that systematic uncertainties in the lens model should not significantly bias an estimate of the \hi mass for these sources, given their expected \hi mass ranges. This may be due to the relatively extended and smooth spatial distribution of the idealised \hi disks, which averages out small-scale variations in the lens models.

Even if the \hi mass can be derived without significant systematic error, it is important to consider whether other galaxy components, such as the stellar mass, may still suffer from systematic magnification biases. Differential magnifications between emission components can impact quantities like $M_\mathrm{HI}/M_\star$ \citep[e.g.][]{deane_2013, spilker_2015}. The stellar or molecular gas components may not have the same extent and regularity as the \hi distribution, which means that small-scale variations in the lens model could exert a greater influence on the magnification. See the right hand panels in Figure~\ref{profiled_sources_srcplane} for maps of the spatial variation at the source locations for the GLAFIC models.

Future simulations could explore the feasibility of extracting quantities such as $M_\mathrm{HI}/M_\star$ or $M_\mathrm{HI}/M_\mathrm{H_2}$ by ray tracing different galaxy components. This would provide a more comprehensive understanding of how gravitational lensing affects the interpretation of multiwavelength galaxy observations.

\section{Conclusion}
We have investigated the potential for measuring the neutral hydrogen content of gravitationally lensed galaxies behind the Hubble Frontier Field Clusters. Towards this aim, we have achieved the following:
\begin{enumerate}
    \item Performed \hi lensing simulations of 401 known galaxies behind the Frontier Field clusters. 
    
    \item Identified several galaxies with both high magnification and predicted high \hi mass at $z\gtrsim 0.7$.
    
    \item Detailed the relationship between source \hi mass and magnification for three of these galaxies, thereby providing a constraint on the \hi flux - \hi mass modelling degeneracy.
    
    \item Computed approximate observing time requirements for the three profiled galaxies, with the MeerKAT radio telescope UHF receivers. Among these, the most promising source was the Abell~370 Great Arc with an estimated observing time requirement of $\tau_{\rm 5 \sigma} = 16^{+35}_{-10}$~hr.

    \item Demonstrated that if the assumptions of the model are fulfilled, given a $5~\sigma$ detection, the reconstructed \hi mass could be constrained within a factor of $\sim 2.5$ for a 95 per cent confidence interval.
    
    \item Found that lens model systematic errors are subdominant to statistical uncertainties for the two profiled galaxies behind Abell~370, and hence should not significantly bias \hi mass measurements in the expected mass ranges.
\end{enumerate}

Our simulations reveal that systems with both high mass and high magnification exist, but are uncommon, within the studied redshift range for these four clusters. Nonetheless, a key next step will be to quantify the occurrence of such systems, considering the recent detection of numerous new group and cluster scale lenses through novel wide-field image surveys \citep[e.g.][]{sonnenfeld_2018,jaelani_2020}. This is particularly relevant in view of using cluster-scale lensing as a high-redshift \hi measurement tool in the Square Kilometre Array era.

\section*{Data Availability}
There are no new data associated with this article.

\bibliographystyle{mnras}
\bibliography{bibliography_processed}


\appendix
\section{Ray tracing}\label{sec:raytracing}
\begin{figure}
	\begin{center}
		\includegraphics[width=0.8\columnwidth]{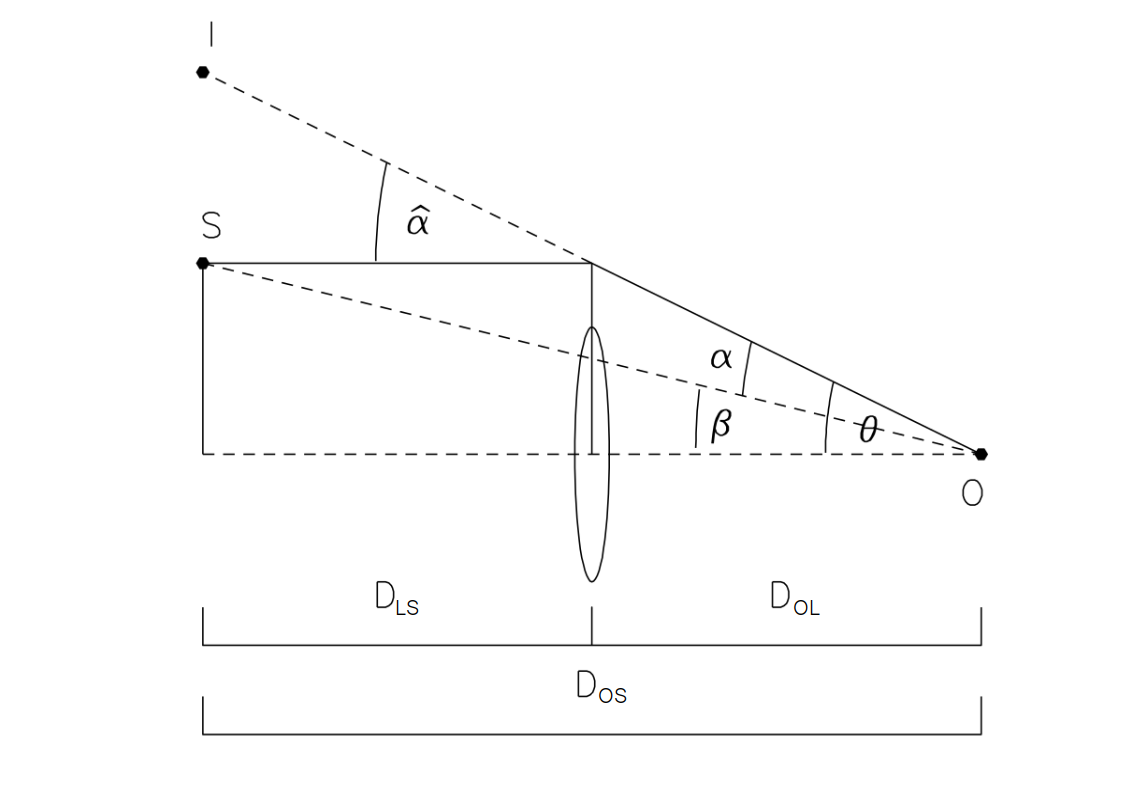}
		\caption{Diagram showing the key aspects of the thin screen lensing geometry in one dimension. The light ray travels from a source, S, at an angle $\beta$ with respect to the observer, O, and is deflected by the lens by an angle $\hat{\alpha}$ such that the image, I, is at an angle $\theta$. The angular diameter distances from observer-to-source, observer-to-lens and lens-to-source are $D_{\rm OS}$, $D_{\rm OL}$ and $D_{\rm LS}$ respectively. The reduced deflection angle $\vec{\alpha}$ is related to the deflection angle $\hat{\alpha}$ through Equation~\ref{eq:reduced_defangle}.\label{fig:lens_geom}}
	\end{center}
\end{figure}

Figure~\ref{fig:lens_geom} illustrates the geometry of the thin screen lens. Following \citet{narayan_1996}, a light ray originates from a source, S, which is situated in a plane (referred to as {\it source}-plane) at a distance $D_{\rm OS}$ and is indexed by an angle $\vec{\beta}$ (with respect to an arbitrary axis which we will choose to be centered on the lens). The ray intersects the image plane which is at a distance $D_{\rm OL}$ and is indexed by angle $\vec{\theta}$. The ray is deflected by an angle $\vec{\hat{\alpha}}(\vec{\theta})$. Certain deflections will result in an image, I, being seen by the observer, O. Note that the distances represented are angular diameter distances and angular diameter distances do not add, $D_{\rm OS}\ne D_{\rm OL} + D_{\rm LS}$, in a non-Euclidean universe.

From this geometry we see that
\begin{equation}
\vec{\theta}D_{\rm OS}=\vec{\beta}D_{\rm OS}+\vec{\hat{\alpha}}D_{\rm LS}\,  
\end{equation}
and therefore,
\begin{align}
	\vec{\theta} &= \vec{\beta} + \frac{D_{\rm LS}}{D_{\rm OS}} \vec{\hat{\alpha}} (\vec{\theta}) \nonumber\\
	& = \vec{\beta} + \vec{\alpha} (\vec{\theta})\ , \label{eq:lens_equation}
\end{align}
which is known as the {\it lens equation}. In  Equation~\ref{eq:lens_equation}, we have defined the reduced deflection angle, 
\begin{equation}
	\vec{\alpha}\equiv\dfrac{D_{\rm LS}}{D_{\rm OS}} \vec{\hat{\alpha}}\, , \label{eq:reduced_defangle}
\end{equation}
where the factor $D_{\rm LS}/D_{\rm OS} \equiv \epsilon$ is known as the lens efficiency.

The lens equation states that given a coordinate in the image plane $\vec{\theta}$, the deflection angle at that point $\vec{\hat{\alpha}} (\vec{\theta})$, and the redshifts of the source and lens, then the corresponding coordinate in the source-plane $\vec{\beta}$ is uniquely determined. It follows from this that surface brightness is conserved in the lens mapping. In addition, if the flux distribution in the source-plane is known then the flux value at any $\vec{\theta}$ can be calculated from the lens equation.

\begin{figure}
	\begin{center}
		\includegraphics[width=\columnwidth]{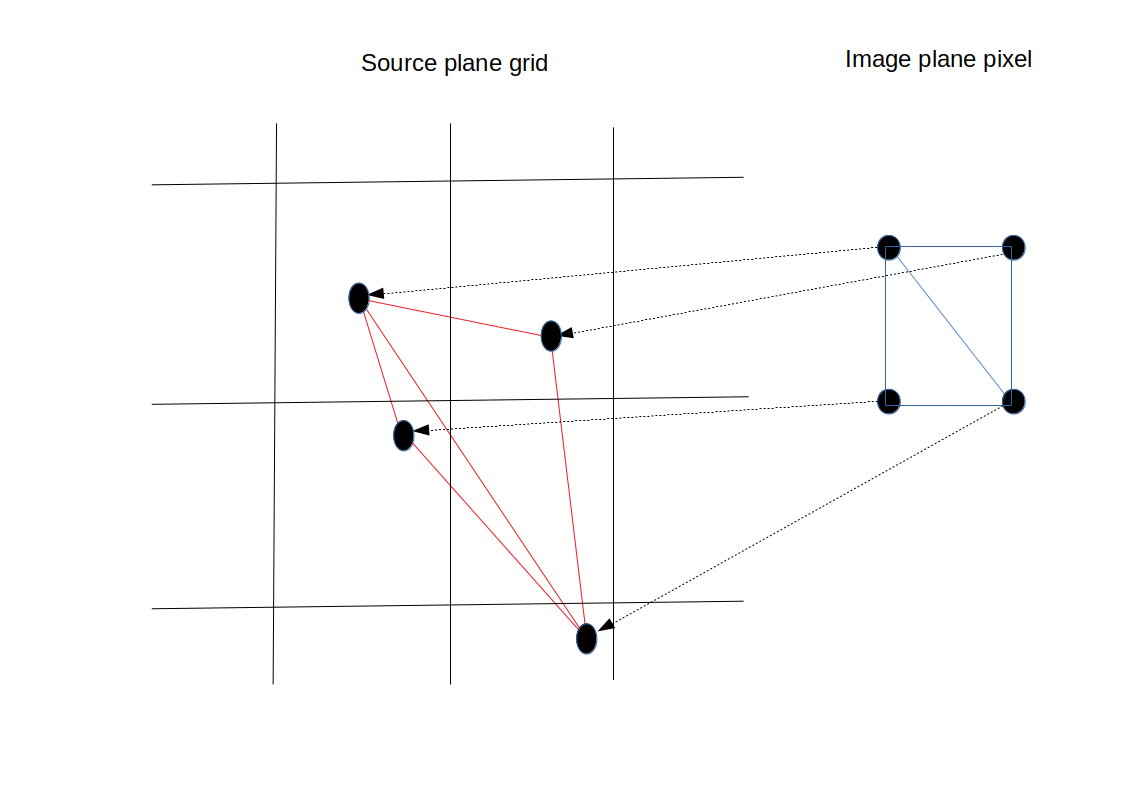}
		\caption{An illustration of a pixel in the image plane (right side) ray-traced to the source-plane (left side). In this case, the deflection angle changes over the image plane pixel which is transformed into an irregular polygon in the source-plane. \label{fig:triangle_ray}}
	\end{center}
\end{figure}

The lens equation allows us to transform point coordinates in the image plane to point coordinates in the source-plane. However, in practice the deflection angle $\vec{\hat{\alpha}} (\vec{\theta})$ is given as a pixelated grid. Consider the case of ray tracing a pixel from the image plane to source-plane via the lens equation. A simple approximation would be to ray trace the coordinate at the centre of the image plane pixel and set its flux value to that of the corresponding source-plane point. To make this approximation more accurate, the source flux distribution can be interpolated to enable sub-pixel resolution. We will refer to this as the `pixel-centre' approximation. There is an inaccuracy associated with purely considering the pixel centroid. As the four corners of a pixel in the image plane actually have slightly different deflection angles, the resulting shape of the pixel in source-plane will be an irregular polygon as shown in Figure~\ref{fig:triangle_ray}.

In the case of point sources, where fine resolution is needed, instead of ray tracing squares, triangles are preferred as the resulting shape is always convex \citep[i.e. another triangle, see Figure~\ref{fig:triangle_ray};][]{keeton_2001}. The value of each triangle in the image plane can then be calculated by averaging together all the pixels in the source-plane which that triangle intersects, where the average is weighted by the area of intersection (see Figure~\ref{fig:triangle_ray}). Although this method is more precise, it is significantly more computationally expensive due to the calculation of the areas of intersection.

\begin{figure}
	\begin{center}
		\includegraphics[width=\columnwidth]{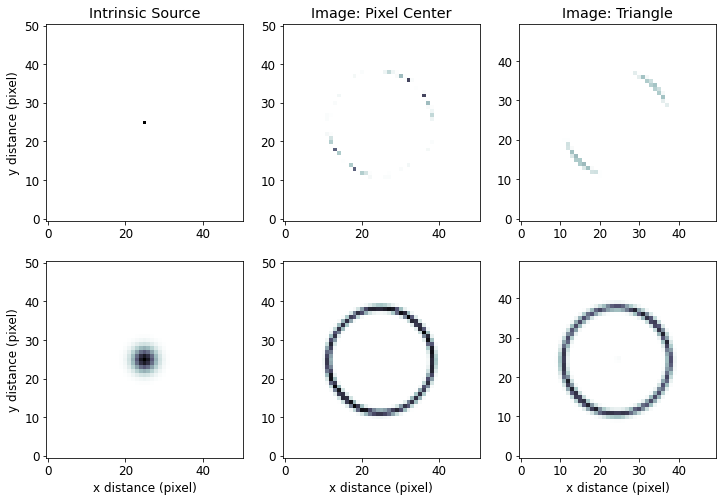}
		\caption{An example of two sources (left column) lensed by a point mass using two different ray tracing schemes: the pixel center approximation (middle column) and the triangle weighting scheme (right column). The percentage difference in magnification between the two methods is approximately 18 per cent for the point source and 8 per cent for the extended source. \label{fig:ray_trace_compare}}
	\end{center}
\end{figure}

In general, when dealing with extended lensed sources, the pixel centre method provides a reasonable approximation and is typically used in the lensing community \citep[Oguri, M, private communication, Nightingale, J, private communication][]{}. An example comparison between the two ray-tracing methods is shown in Figure~\ref{fig:ray_trace_compare}, where the percentage difference in the magnification between the two models decreases from 18 per cent to 8 per cent for  point and extended sources respectively. As \hi is extended and diffuse, we will make use of the pixel centre ray-tracing method in this work.

\section*{Acknowledgements}
We thank Masamune Oguri and James Nightingale for discussions on ray tracing. This research was supported by the South African Radio Astronomy Observatory, which is a facility of the National Research Foundation, an agency of the Department of Science and Technology. RPD's research is funded by the South African Research Chairs Initiative of the DSI/NRF. DO is a recipient of an Australian Research Council Future Fellowship (FT190100083) funded by the Australian Government. This work utilizes gravitational lensing models produced by PIs Bradač, Natarajan \& Kneib (CATS), Merten \& Zitrin, Sharon, Williams, Keeton, Bernstein and Diego, and the GLAFIC group. This lens modeling was partially funded by the HST Frontier Fields program conducted by STScI. STScI is operated by the Association of Universities for Research in Astronomy, Inc. under NASA contract NAS 5-26555. The lens models were obtained from the Mikulski Archive for Space Telescopes (MAST).

\bsp	
\label{lastpage}
\end{document}